# Particle size effect on strength, failure and shock behavior in Polytetrafluoroethylene-Al-W granular composites

E.B. Herbold, V.F. Nesterenko, D.J. Benson

*Department of Mechanical and Aerospace Engineering, University of California at San Diego, La Jolla, California 92093-0411, USA,*

J. Cai

*Materials Science and Engineering Program, University of California at San Diego, La Jolla, California 92093-0418, USA*

K.S. Vecchio, F. Jiang

*Department of NanoEngineering, University of California at San Diego, La Jolla, California 92093-0418, USA*

J. W. Addiss, S. M. Walley, W. G. Proud

*Cavendish Laboratory, Madingley Road, Cambridge, CB3 0HE, UK*

(June 9, 2008)

The variation of metallic particle size and sample porosity significantly alters the dynamic mechanical properties of high density granular composites processed using a cold isostatically pressed mixture of polytetrafluoroethylene (PTFE), aluminum (Al) and tungsten (W) powders. Quasi-static and dynamic experiments are performed with identical constituent mass fractions with variations in the size of the W particles and pressing conditions. The relatively weak polymer matrix allows the strength and fracture modes of this material to be governed by the granular type behavior of agglomerated metal particles. A higher ultimate compressive strength was observed in relatively high porosity samples with small W particles compared to those with coarse W particles in all experiments. Mesoscale granular force chains comprised of the metallic particles explain this unusual phenomenon as observed in a hydrocode simulation of a drop-weight test. Macrocracks forming below the critical failure strain for the matrix and unusual behavior due to a competition between densification and fracture in dynamic tests of porous samples were also observed. Shock loading of this granular composite resulted in higher fraction of total internal energy deposition in the soft PTFE matrix, specifically thermal energy, which can be tailored by the W particle size distribution.

**INTRODUCTION**

Mixtures containing polytetrafluoroethylene (PTFE) and aluminum (Al) are known to be energetic under dynamic and/or thermal loading.[1-6] These materials generate a large quantity of heat during the reaction driven by bulk mechanical deformation as in thermites.[7] This paper considers the mechanical behavior of PTFE-Al-W composites where tungsten (W) was used to increase density[8] keeping the composition of PTFE-Al mixture close to the stoichiometric value.

Tailoring the mechanical and chemical properties of reactive materials is important for various applications. For example, varying particle size and morphology in pressed explosives[9-12] or layer thicknesses in laminates[13] significantly alter the shock-sensitivity and the rate of energy release. Ignition sites within composite energetic materials under compressive load have been ascribed to the stress or force-chain formation in granular energetic materials.[14-15]

It should be emphasized that under conditions of shock wave loading of the mixture PTFE-Al-W large, high strength heavy particles (W in this case) may transform their kinetic energy, which they had before the shock front, into the thermal energy of small, soft particles behind the shock front.[16,17] Thus, this mixture under shock loading naturally transforms the kinetic energy of one component (W) into the internal energy of another, in this case of PTFE and Al particles, thus promoting the ignition of the chemical reaction between them.

One of the distinguishing features of reactive materials is that the stored chemical potential is released upon material deformation and is usually not self-propagating in contrast to energetic materials, which may support detonation waves depending on the initiation energy threshold. The material structure may be optimized at the meso-scale in order to facilitate the desirable release of mechanical and subsequent chemical energy in hot spots (e.g. shear bands) to optimize the amount and/or rate of energy release by material.

In this paper the overall strength, modes of dynamic failure and behavior under shock loading of PTFE-Al-W granular composites are the primary points of investigation. The experimental samples were prepared to investigate how the first two aspects are affected by varying the metallic particle size, which also determines the inherent porosity for a given densification pressure. Quasi-static, Hopkinson bar, and drop-weight experiments were performed and multi-material Eulerian hydrocode simulations of drop-weight tests with the different types of samples are used to explain the observed experimental results. The behavior of these granular composites under shock loading are also examined numerically.

**EXPERIMENTAL**

**Sample preparation**

The three main types of specimens had identical mass ratios and were prepared by using cold isostatic pressing (CIPing) the powder mixtures (17.5 % PTFE, 5.5 % Al, and 77 % W, by weight). The density and porosity of the two the samples with large W particles varied only in the pressing conditions (see Table I). Under the same CIPing conditions (pressing pressure, time and specimen size) the density of porous PTFE-Al-W with fine W particles was 6 g/cm$^3$ while the density of dense PTFE-Al-W with coarse W particles was 7.1 g/cm$^3$, which is close to the theoretical maximum density. The level of porosity in the sample with fine W particles indicates that it resists densification in comparison with the sample with large W particles. At the same pressing condition, the mixture of PTFE and Al powders can be fully densified.[18] It is important to emphasize that although the dense sample with coarse W particles is relatively free of pores and cracks, the average spacing between metallic grains is expected to be greater than in the sample with fine W particles with the same constituent volume ratios.

The metallic powder granules had an approximately spherical shape with the following sizes: 2 μm Al powder (Valimet H-2), coarse W powder sieved to less than 44 μm (Teledyne, -325 mesh) and fine W powder sieved to less than 1μm (Alfa-Aesar). The size of the PTFE powder grains was approximately 100 nm (DuPont, PTFE 9002-84-0, type MP 1500J). The mixed powders were ball milled in an SPEX 800 mill for 2-10 minutes using alumina balls to reduce agglomeration. Typical dimensions of the CIPed cylindrical specimens were 10 mm high and 10.44 mm in diameter and masses equal to 4 to 5 grams. Three to six samples of each type of composite were tested in each type of experiment.

TABLE I: Properties of various specimens.

|  | Dense PTFE-Al-coarse W | Porous PTFE-Al-fine W | Porous PTFE-Al-coarse W | Pure Dense PTFE |
|---|---|---|---|---|
| Size of W Particles (μm) | <44 | <1 | <44 | … |
| CIPing Pressure (MPa) | 350 | 350 | 20 | 350 |
| Density (g/cm$^3$) | 7.1±0.4 | 6.0±0.3 | 6.0±0.3 | 2.1±0.1 |
| Porosity (%) | 1.6 | 14.3 | 14.3 | 4.5 |

Porous PTFE-Al-W specimens containing coarse W particles (denoted "porous PTFE-Al-coarse W") were processed with a significantly reduced CIPing pressure (20 MPa) to investigate the behavior of materials with different porosity and different particle sizes of W powder. This resulted in a similar porosity to samples with fine W particles (compare the Porous PTFE-Al-fine W sample with the Porous PTFE-Al-coarse W sample in Table. I). Specimens of CIPed PTFE with a density of 2.1 g/cm$^3$ were also manufactured for the measurements of properties of the PTFE matrix used in the numerical analysis of the composite behaviors.

**Quasistatic tests**

Quasi-static compression tests were performed using the SATEC™ Universal Materials Testing Machine (Instron; Canton, MA). The load was applied at a nominal deformation strain rate of 0.001 s$^{-1}$ and was terminated when the sample was fractured. The ultimate compressive engineering stresses for the composite and densified PTFE samples are listed in Table II.

TABLE II: Quasi-static test results for each specimen.

|  | Dense PTFE-Al-W (coarse W) | Porous PTFE-Al-W (fine W) | Porous PTFE-Al-W (coarse W) | Pure Dense PTFE |
|---|---|---|---|---|
| Quasi-static tests ($10^{-3}$ s$^{-1}$) | 18±1 MPa | 22±6 MPa | 5.8±0.2 MPa | 2.3±0.3 MPa |

In Fig. 1 the modes of failure are compared between the samples. Both shear and axial cracks were observed in the porous samples with small W particles (see Fig. 1 (a) and (b)) and the porous samples with large W particles (see Fig. 1 (c) and (d)). The higher density samples with coarse W powder show failure patterns in a unique combination of axial and circumferential cracks (Fig. 1 (e) and (f)).

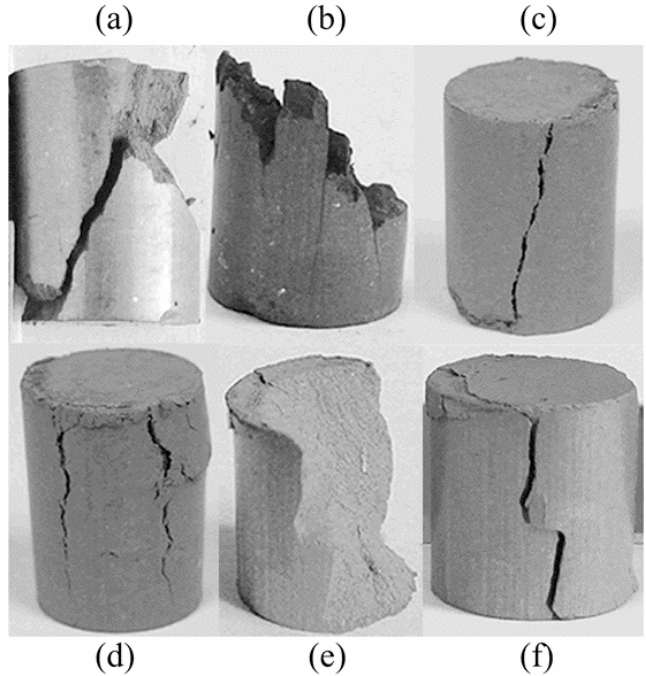

FIG. 1: Fracture detail of various samples after quasi-static testing. (a) Shear crack and (b) axial and shear cracks in the porous fine PTFE-Al-W composite sample; (c) Axial/shear and (d) axial cracks in the porous coarse PTFE-Al-W composite sample; (e), (f) kinked axial/shear cracks in the dense coarse PTFE-Al-W composite sample. All specimens had identical initial dimensions of 10.44 mm diameter and 10 mm height.

Figure 2 shows typical stress strain curves for the three different types of samples shown in Fig. 1. These tests consistently demonstrated that porous PTFE-Al-W composites with fine W particles exhibit the highest compressive strength (see curve (1)). It is natural to expect that porous materials in compression tests fail due to axial cracks caused by tensile stress concentration at the vicinity of pores.[19] However, it is interesting that the observed fracture pattern in the porous PTFE-Al-W containing fine W particles consisted of mainly shear cracks (see Fig. 1 (a) and (b)). It is also important to mention that the sample strength is proportional to the slope in the stress strain relation indicating a the effective modulus at the initial stage of deformation.

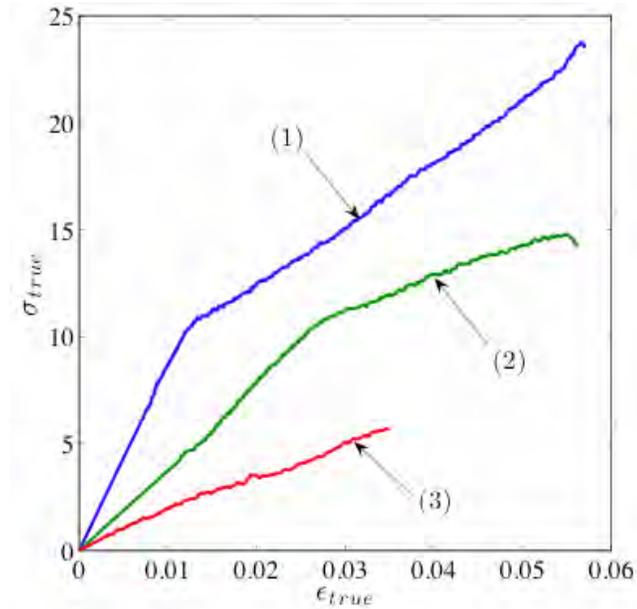

FIG. 2: Quasi-static stress-strain curve of PTFE-W-Al composites with variation of density (dense versus porous composites with coarse W particles) and particle size of W (porous composites with fine W particles versus porous coarse composites). Curve (1) shows the results of the sample with fine W particles and porosity above 14%. Curve (2) shows the results of the sample with large W particles, which was almost fully densified. Curve (3) shows the results of the sample with large W particles CIPed at a lower pressure to induce about 14% porosity.

**Dynamic Hopkinson bar tests**

Dynamic testing was performed using a Hopkinson bar which comprises three 19 mm diameter bars: a 457 mm long C350 maraging steel striker bar, a 1828 mm long C350 maraging steel incident bar and a 1828 mm long magnesium alloy transmitting bar.

Since the investigated materials are generally of lower strength, a linear elastic, low modulus magnesium alloy transmission bar was used to obtain a high signal to noise ratio in the transmitted signals. The Hopkinson Bar testing usually generated only 5% strain in the specimens for the investigated materials. Specimens of CIPed monolithic PTFE were also tested (see Table III) to obtain the ultimate compressive strength of the PTFE matrix and the failure strain (about 0.05). The equations for calculating stress, transmitted ($\varepsilon_T$) and reflected ($\varepsilon_R$) strains and strain rate, in the specimen in the Hopkinson bar experiment are,

$$\sigma(t) = \frac{E_{MS} + E_M}{2} \cdot \frac{A_O}{A} \cdot \varepsilon_T(t)$$
$$\varepsilon(t) = \frac{1}{L} \int_0^t [C_{MS}(\varepsilon_T(t) - 2\varepsilon_R(t)) - C_M \varepsilon_T(t)] dt \qquad (1)$$
$$\dot{\varepsilon}(t) = \frac{C_{MS}(\varepsilon_T(t) - 2\varepsilon_R(t)) - C_M \varepsilon_T(t)}{L}.$$

In Eqs. (1), $C_{MS}$ and $C_M$ denote the sound speed in the maraging steel incident bar (5000 m/s) and the magnesium transmitted bar (5000 m/s) and $A_0$, $A$ denote areas of the bars and the sample with length $L$. The constants $E_{MS}$ and $E_M$ denote the elastic moduli of the maraging steel ($E_{MS}$ = 200 GPa) and the magnesium bar ($E_M$ = 45 GPa).

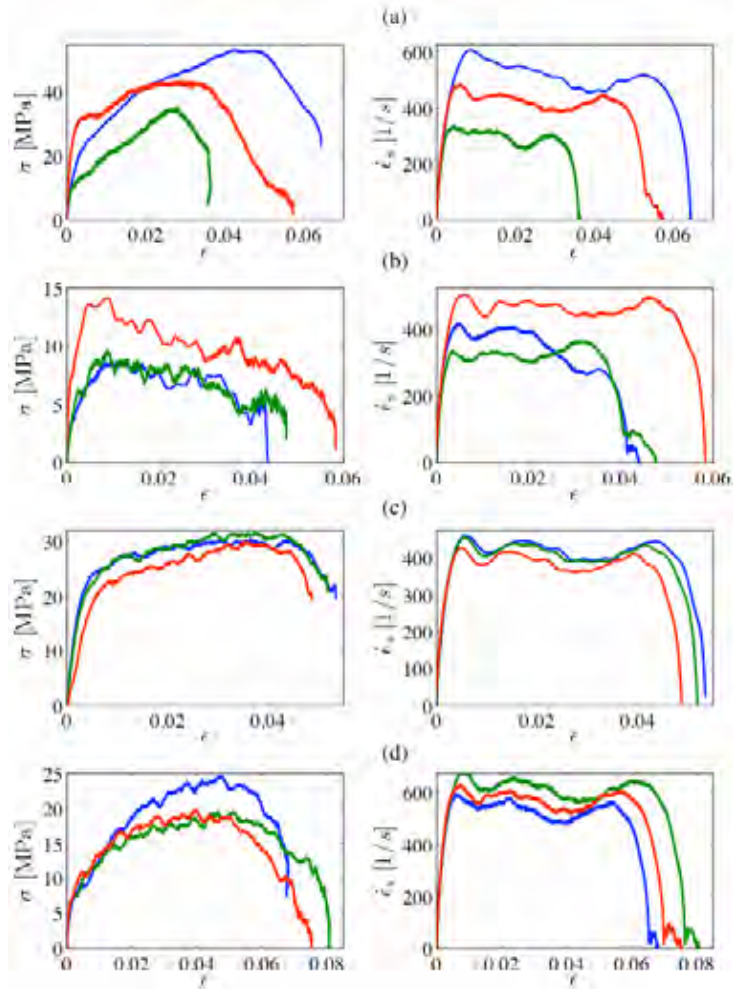

FIG. 3: Hopkinson bar stress-strain and strain-rate-strain curves of a (a) porous PTFE-Al-W composite specimen containing fine W particles, (b) porous PTFE-Al-W composite samples containing coarse W particles, (c) dense PTFE-Al-W composite samples containing coarse W particles and (d) cold isostatically pressed PTFE samples.

The Hopkinson bar results for the three composite and the pure PTFE samples are shown in Fig. 3 and listed in Table III. In Fig. 3 (a) three stress versus strain and strain rate versus strain plots are shown for the porous sample with small W particles. The striking differences between the three experiments may be attributed to the samples ultimate strength, the final strain value and the strong dependence on the *in situ* state of the metallic particles. In other words, the mesostructure of the metallic particles may or may not be conducive to resisting a load at any point during dynamic deformation. However, compared to the other samples, the ultimate strength of this sample is the highest.

TABLE III: Hopkinson bar tests results for each specimen

|  | Dense PTFE-Al-W (coarse W) | Porous PTFE-Al-W (fine W) | Porous PTFE-Al-W (coarse W) | Pure Dense PTFE |
|---|---|---|---|---|
| Hopkinson bar ($500 \text{ s}^{-1}$) | 30±1 MPa | 45±7 MPa | 10±3 MPa | 21±3 MPa |

There is also noticeable variation in the results for the porous sample with large W particles shown in Fig. 3 (b). For example, in one of the tests shown in Fig. 3 (b), the sample exhibits a relatively higher strength throughout the whole range of strain. This may be due to the gradual densification of the sample, which is possible when the CIPing pressure is too low as was also observed in drop weight tests presented later. Even with densification, the ultimate strength of this sample is lower compared to the pure PTFE sample shown in Fig. 3 (d).

The most consistent tests result from the dense sample with large W particles (Fig. 3 (c)). Compared to the Fig. 3 (b), this sample may not easily densify any further during the compression test. Also, the metallic particles are sparsely distributed throughout the PTFE matrix with respect to the porous sample with small W particles (Fig. 3 (a)). This may produce a cushioning effect between the metallic particles and explain the lower average strength of the sample with large W particles in comparison with the sample with small W particles.

**Dynamic "soft" drop weight tests**

The investigated samples have relatively low strength as evident form Tables II and III. The signal in standard drop weight test is obscured by inevitable noise due to oscillations in the drop weight device. A "soft" drop-weight test[8,20] was developed to allow effective testing of low strength specimens. A 201 nitrile O-ring was placed on the top of the upper anvil of the drop-weight apparatus to reduce the mechanical oscillations in the system caused by the impact of the mass. This approach effectively reduced the high amplitude parasitic oscillations. Displacements accumulated in the O-ring can be taken into account in calculation of engineering strains based on separate experiments with another O-ring used as a sample. The results of the dynamic measurements of ultimate engineering compressive stresses are presented in Table IV. Comparing the strengths of the porous PTFE-Al-fine W and the porous PTFE-Al-coarse W, it is clear that porosity itself does not contribute to the higher strength of the porous composite filled with fine W particles.

TABLE IV: Drop-weight tests results for each specimen

|  | Dense PTFE-Al-W (coarse W) | Porous PTFE-Al-W (fine W) | Porous PTFE-Al-W (coarse W) |
|---|---|---|---|
| Drop-weight tests (300 s$^{-1}$) | 32±2 | 55±6 | 11±2 (40±11) |

The porous samples with coarse W particles also demonstrated an unusual behavior as in Hopkinson bar tests. Figure 4 shows the engineering stress for each sample plotted versus time. In each case a copper ring surrounded the sample to arrest the deformation at different strains in the vertical direction. In Fig. 4 (a) some of the porous samples with large W particles exhibited a very low strength and failed in shear at approximately 11

MPa (curve 1) while others exhibited considerably higher strength above 50 MPa (curve 2). The higher ultimate compressive strength may be attributed to the gradual densification during the initial stage of deformation, which leads to a considerably increased strength relative to those samples that appear to fail almost immediately upon impact.

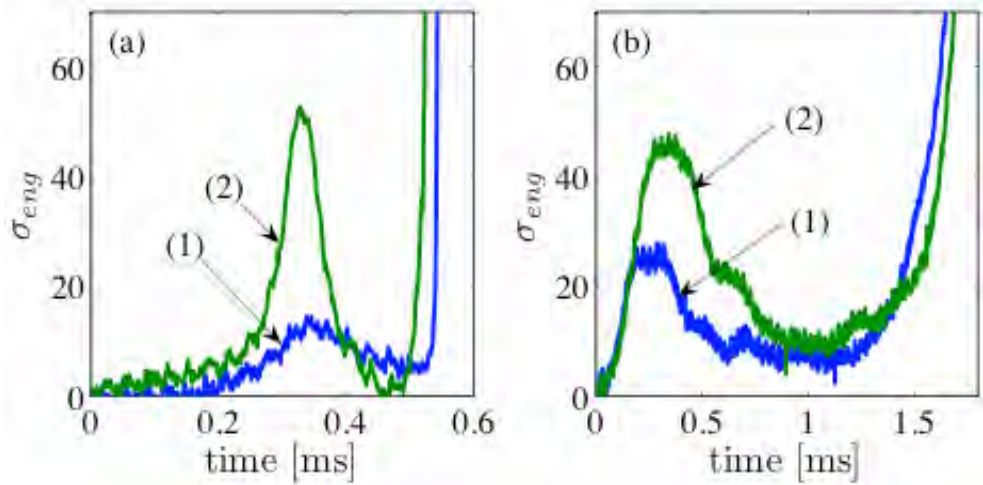

FIG. 4: Stress vs. time curves obtained in drop weight experiments. (a) Curves (1) and (2) correspond to the porous sample with coarse W particles. The remarkable difference between the two curves is due to the densification of the sample before fracture. (b) Curve (1) corresponds to the densified sample with coarse W particles. Curve (2) corresponds to the porous sample with fine W particles.

The ultimate compressive strength of the *in situ* densified samples is greater than the pressing pressure (20 MPa) and comparable to that of the dense samples with coarse W. Such behavior can be expected when the ultimate compressive strength is comparable to the densification pressure used to prepare samples and was not observed for porous samples with fine metallic particles or denser samples with coarse W particles (Table IV).

There is clearly some competition occurring between the compaction of the soft visco-elastic matrix and fracture during the deformation process as evidenced by some samples failing by shear at low strains while others are *in-situ* densified during the test leading to an increased strength (compare curves (1) and (2) in Fig. 4).

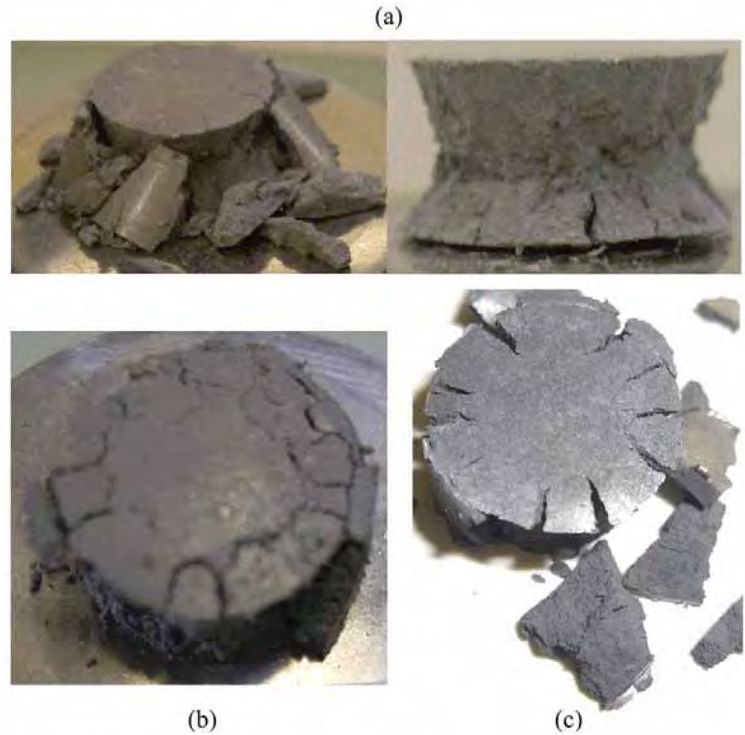

FIG. 5: (a) Porous sample (sample 73, engineering strain 0.15) with lower strength containing coarse W particles post drop-weight test corresponding to curve (1) in Fig. 4 (a). (b) Porous sample (sample 65), engineering strain 0.26 containing coarse W particles corresponding to curve (2) in Fig. 4 (a). (c) Porous sample (sample 64), engineering strain 0.25 with fine W particles corresponding to curve (2) in Fig. 4 (b).

Figure 5 shows recovered samples where the deformation was interrupted at corresponding engineering strains. The stress-strain curve associated with the sample shown in Fig. 5 (a) is similar to curve 1 shown in Fig. 4. The sample has failed by shear localization. A symmetrical pattern of cracks can be seen in the fractured samples (Fig. 5), though the sample's microstructure is inherently random.

Debonding of the metal particles from the matrix and the fracture of matrix were two major mesoscale mechanisms for the failure of the specimen.[21] The relevant mesoscale mechanism of shear localization at low levels of strain is considered in Ref. 22 and examples of shear localization due to micro-fracture mechanisms leading to reaction in granular materials (for example, in granular $Al_2O_3$ and solid and granular SiC) can be found in Refs. 17,23,24.

## NUMERICAL MODELING AND DISCUSSION

### Numerical modeling of dynamic drop weight test

Results of quasistatic and dynamic tests are consistent with each other with respect to the influence of particle size on sample strength. Drop weight tests were selected for numerical analysis due to well defined boundary conditions determined by the approximately constant velocity of the dropped weight. A two-dimensional numerical simulation was used to explain the unusual dependence of increased ultimate compressive strength with the decreased of size of metallic particles (high porosity, low density samples) in drop-weight tests. Two dimensional granular packings were used for modeling of shock compaction and determination of microkinetic energy[25] as photoelastic discs have been used effectively as analogs of energetic materials in the literature.[11,12] The two dimensional metallic particle mesostructure of the composites do not reproduce the coordination number as in three dimensional packings, but numerical modeling can help to estimate the level of influence the force chains have on the global behavior of the samples.

Two samples using a randomly distributed mixture of fine (1 $\mu$m) W and Al particles (2 $\mu$m) (sample 1, Fig. 6 (a)) and coarse (10 $\mu$m) W and Al particles (2 $\mu$m) (sample 2, Fig. 6 (b)) of circular shape are used in finite element calculations to investigate the force chain effect. The weight and 'volume' fractions of each the sample constituents were similar in the calculations and experiments (e.g. mass fractions: 20% PTFE, 6% Al, 74% W, and 'volume' fractions: 59% PTFE, 13% Al, 28% W).[26] In numerical calculations a samples contained about 52 particles of Al, 406 of small W particles or 4 large W particles, correspondingly. In experiments samples contained about around 50 billion Al particles and 900 billion small W particles or 85000 large particles respectively. It is interesting that despite that many orders difference in number of particles the considered representative volume of granular composite with arrangements shown in Fig. 6 reflect qualitative difference in behavior detected in experiments, as we will show.

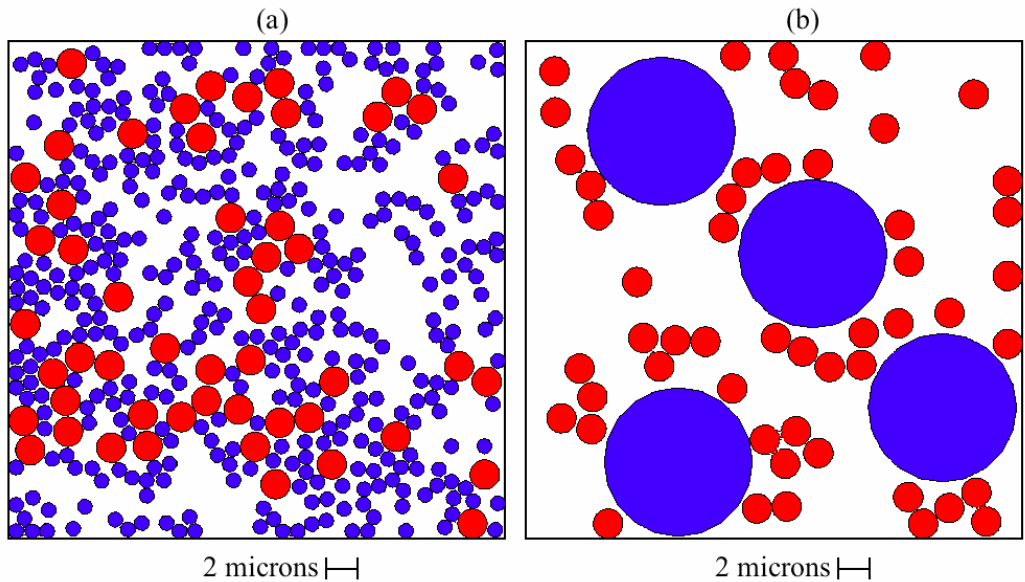

FIG. 6: (a) PTFE-W-Al specimen ('sample 1') using 2 $\mu$m Al particles and 1$\mu$m W particles. (b) PTFE-W-Al specimen ('sample 2') using 2$\mu$m diameter Al particles and 10$\mu$m diameter W particles.

A two-dimensional Eulerian Hydrocode[27] is implemented to simulate the behavior of the sample at high strain rates in drop weight tests. Each material in the mixture has a different equation of state, physical and mechanical properties. The material model used for PTFE was the Johnson-Cook with Failure[28]

$$\sigma_y = \left[A + B(\bar{\varepsilon}^p)^n\right]\left[1 + C \ln \dot{\varepsilon}^*\right]\left[1 - T^{*m}\right], \tag{4}$$

where $A = 11$ MPa, $B = 44$ MPa, $n = 1$, $C = 0.12$ and $m = 1$ (determined from the data in Ref. 29). The material failure criteria was based on the equation,

$$\varepsilon_f = \left[D_1 + D_2 \exp(D_3 \sigma^*)\right]\left[1 + D_4 \ln \dot{\varepsilon}^*\right]\left[1 + D_5 T^*\right], \tag{9}$$

where $D_1 = 0.05$ was obtained from quasi-static and Hopkinson bar experimental data of pure CIPed PTFE samples and the $D_2$, $D_3$, $D_4$ and $D_5$ were set equal to zero as a first approximation. The Gruneisen form of the equation of state was used to define the pressure in compression and tension in PTFE with parameters presented in Ref. 30.

The Johnson-Cook material model without failure was used for the W and Al particles. The Johnson-Cook parameters for W are $A = 1.51$ GPa, $B = 177$ MPa, $n = 0.12$, $C = 0.016$, and $m = 1$. The Johnson-Cook parameters for Al are $A = 265$ MPa, $B = 426$ MPa, $n = 0.34$, $C = 0.015$, and $m = 1$.[31] Since the particle deformation or fracture during dynamic loading is minimal in the mixture with PTFE the equation of state used for W and Al was linear elasticity with the bulk and shear moduli equal to $K = 300$ GPa and $G = 160$ GPa for W and $K = 76$ GPa and $G = 27.1$ GPa for Al.

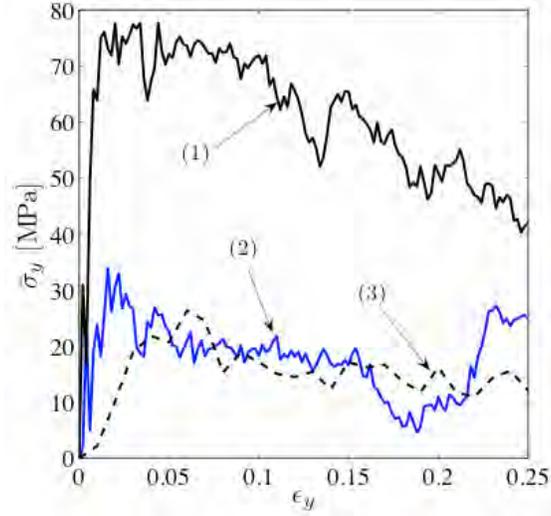

FIG. 7: Average engineering stress at the top of the sample plotted against the 'global' strain for a sample using small W particles (sample 1, curve 1) and a sample using large W particles (sample 2, curve 2). Note the stress increases in curve 1 after 0.13 global strain while the curve 2 coincides with the results for pure CIPed PTFE (curve 3).

The curves in Fig. 7 show that the first compressive stress maxima of sample 1 (curve 1) is 78 MPa; and the corresponding stress of sample 2 at 35 MPa (curve 2) is significantly lower. The level of maximum stress for sample 2 is close to the experimental data (Table IV), but is significantly larger for sample 1. The small number of metal particles and their specific configuration used in the calculations may be responsible for this difference. The results for a pure, densified PTFE are shown in Fig. 7, curve 3 for comparison. Three dimensional numerical calculations with larger number of particles and higher coordination number may provide better agreement with experiments.

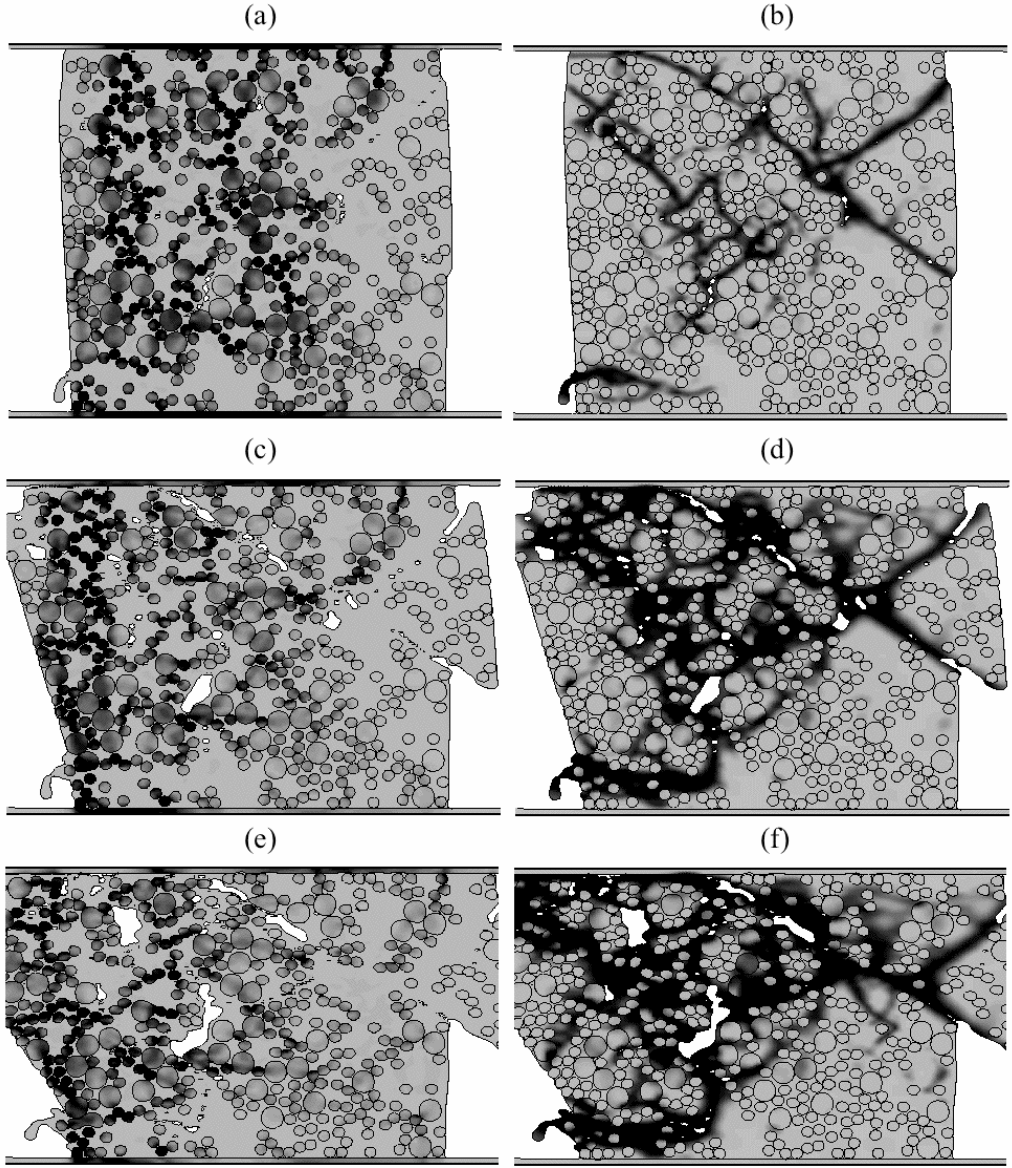

FIG. 8: The color intensity varies from light gray (0 MPa) to dark gray (≥500 MPa) for the von Mises stress and 0 to ≥ 0.05 plastic strain in 'sample 1'. The von Mises true stress distribution (a) and local effective plastic strain (b) at 0.022 'global' strain. The von Mises true stress distribution (c) and local effective plastic strain (d) at 0.042 'global' strain. The von Mises true stress distribution (e) and local effective plastic strain (f) at 0.238 'global' strain.

The von Mises stress and equivalent plastic strain distributions for the sample with fine W particles (sample 1) are shown in Fig. 8 at 0.022, 0.042 and 0.238 'global' strain. Several force chains are apparent starting from the top through the bottom of the sample. This can be compared to the sudden increase in the stress-strain plot shown in curve 1 in Fig. 7 at the corresponding strain. Upon further deformation, this force chain disintegrates and a macrocrack starts in the matrix at a global strain less than the critical failure strain of matrix material (0.05, from Hopkinson bar experiments), resulting in the decrease in stress in curve 1, Fig. 7. Thus, the maximum global stress attained in compression may occur after the onset of failure in the matrix.

Force chains propagating through the cracks in the matrix can be reactivated upon further deformation.[8,31] This self-organization of metallic particles was accompanied by a macro-crack, which is in qualitative agreement with the observed failure in experiments. The local effective plastic strain in the sample above this crack shows that the damage in the PTFE matrix is distributed around the metal particles. It is important to maintain the damage throughout the bulk of the sample to enhance a possible chemical reaction between PTFE and Al.

Several features can be observed from the results of the calculation relating to the first sample. The vertical and horizontal displacements of the metallic particles, initiated by the vertical displacement of the top boundary, are comparable to their sizes resulting in force chains being created, destroyed and reactivated (with different particles) in the course of sample deformation and fracture. It is interesting that the progressive local fracture of the PTFE matrix corresponds to the spikes of global stress (compare curve 1 in Fig. 7 with Fig. 8 (a) and (b)) and the global stress in the highest peak is observed in the heavily fractured sample (compare curve 1 in Fig. 7 with Fig. 8 (e) and (f)). This is due to the disintegration of the matrix being accompanied by a local dense packing of metal particles resisting further deformation.

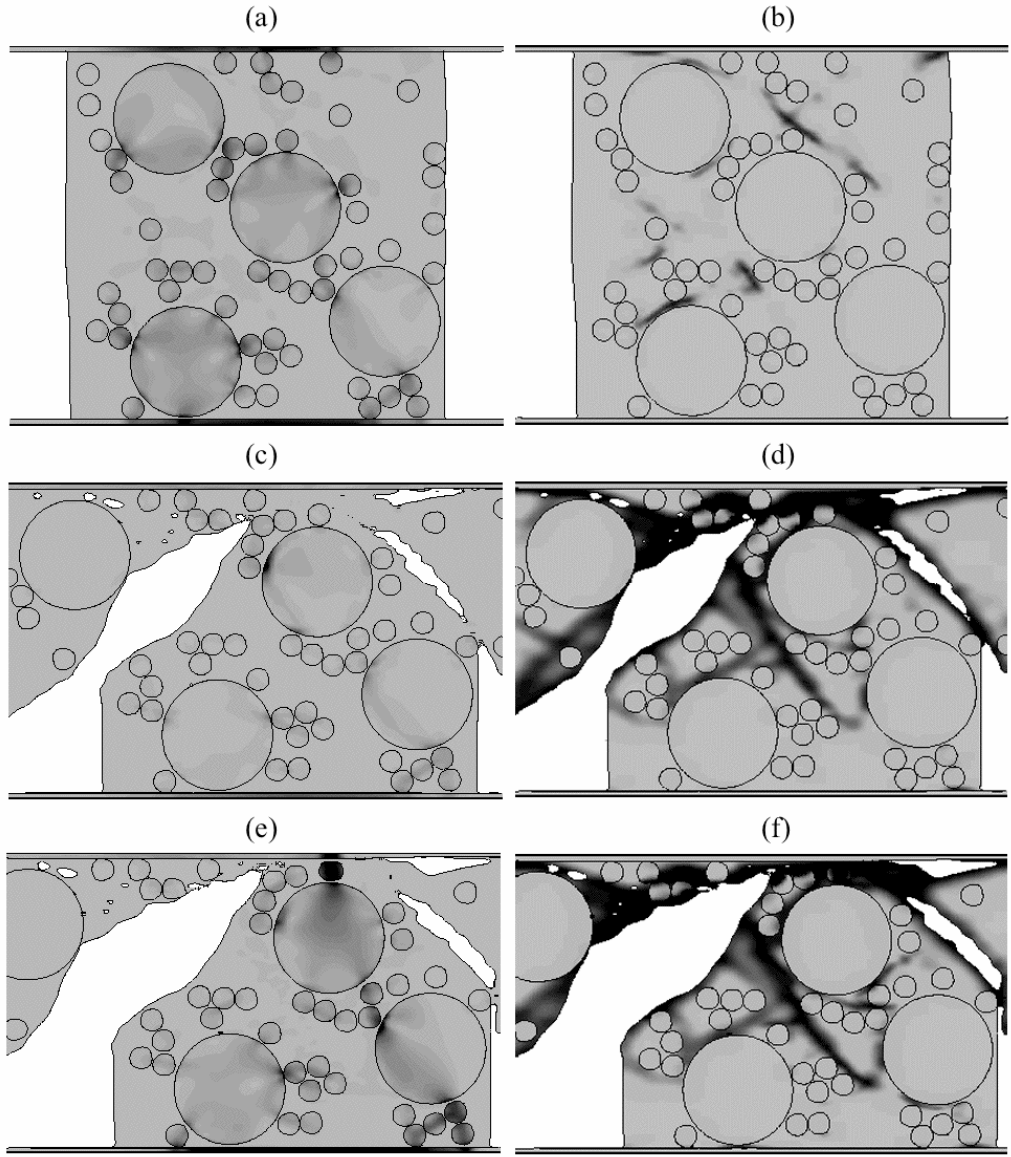

FIG. 9: The color intensity varies from light gray (0 MPa) to dark gray (≥500 MPa) for the von Mises stress and 0 to ≥ 0.05 plastic strain in 'sample 2'. The von Mises true stress distribution (a) and local effective plastic strain (b) at 0.014 'global' strain. The von Mises true stress distribution (c) and local effective plastic strain (d) at 0.186 'global' strain. The von Mises true stress distribution (e) and local effective plastic strain (f) at 0.23 'global' strain.

The second sample (see Fig. 6 (b)) did not have a particle distribution conducive to force chain activation. 'Through-thickness' force chains are not activated as strongly as in the first sample (Fig. 6 (a)) up to 0.25 global strain, though groups of particles have created localized chains. The macro-cracks formed in this sample (Fig. 9) prohibited any bulk-distributed damage. Separate calculations (Fig. 7 curve 3) with a pure PTFE sample have a stress strain behavior very similar to curve 2 in Fig. 7. This suggests that only the matrix material resisted the load.

The vertical and horizontal displacements of the metal particles in the second sample are also comparable to their sizes and to the size of the sample. The calculations also demonstrated that the stress spikes were not related to the activation of force chains propagating from top to bottom, though shorter force chains were activated. The metal particles also initiated shear macro-cracks in the PTFE matrix propagating at 45 degrees (Fig. 9 (c)-(f)) from the direction of compression similar to the behavior observed in experiments (see Fig. 1 (c), (e) and (f) and Fig. 5 (a) and (b)).

The two-dimensional calculations presented demonstrate that force chains created by circular metallic particles are a probable cause of the higher strength of these mixtures with volume content similar to three-dimensional packing of spherical metallic particles in experiments. It should be emphasized that the dependence of the ultimate sample strength on particle size should be sensitive to particle morphology (for example, particles elongated in one direction can exhibit a different dependence of strength on particle size).

The results presented indicate that the skeleton of small metal particles can significantly affect the strength of this granular composite. At the same time the specific distribution of a three dimensional network of force chains can be more complex than that presented in our two dimensional calculations. For example, they may be created by the contacts of agglomerated fine particles as schematically shown for example in Fig. 10 (b). The same volume of metal particles, when a fraction consists of coarse particles, is less conducive to creating a skeleton that is able to affect the global strength of granular composite (Fig. 10 (a)). The correlation of higher strength and stiffness (Fig. 2), observed in experiments, suggests that increase of composite strength is not related to the expected better bonding between agglomerated fine particles and matrix. Numerical calculations demonstrated that forces responsible for particle agglomeration do not significantly contribute to the strength of the granular composite.

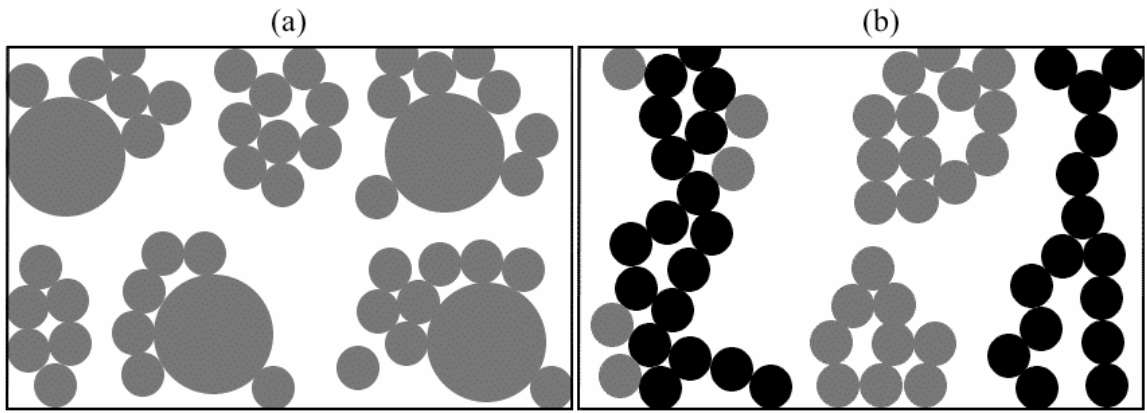

FIG. 10: Different types of metallic particle agglomerate distributions within the soft PTFE matrix (not to scale). (a) Groups of metallic particles coalesce but do not interact immediately upon loading. The interaction of these groups does not contribute to the effective elastic modulus, but may contribute to the critical failure stress as these groups interact with one another during compression test. (b) The metallic particle force chains are highlighted by a darker color for distinction.

These groups immediately interact with one another since their interaction was created during the CIPing process (which resulted in arrested compaction and porosity). The alignment of these groups of particles augments the effective elastic modulus as well as the critical failure strength of the sample. This is also supported by comparing the volume fraction of metallic particles in CIPed mixtures with PTFE and the volume fraction of them at tapped densities of mixtures of coarse W and fine Al and fine W and fine Al powders taken at the same mass ratio as in the mixture.

TABLE V. Density and volume fraction of solid components in tapped powders

| Powder | Theoretical density of solid (g/cm$^3$) | Tapped density of powder (g/cm$^3$) | Ratio % (powder: solid) |
|---|---|---|---|
| Coarse W | 19.3 | 50.2/5.3=9.5 | 49 |
| Fine W | 19.3 | 30/7.2=4.2 | 22 |
| 2-µm Al | 2.7 | 5.7/5.3=1.1 | 40 |
| Coarse W: 2-µm Al 77wt%: 5.5wt% | 13.7 | 53.1/5.6=9.5 | 69 |
| Fine W: 2-µm Al 77wt%: 5.5wt% | 13.7 | 32.2/8.6=3.7 | 27 |

Separate packing studies of the metallic powders were performed using the same constituent mass fractions as in the samples with PTFE. The results are given in Table V. The combined volume fraction of fine W and fine Al particles in the mixture with PTFE for a CIPed sample with density 6 g/cm$^3$ is 0.36. This is higher than the volume fraction of the granules (0.27) in the tapped mixture of fine W and fine Al taken at the same mass ratio as in the mixture with PTFE. This means that the force chains supporting the mesostructure in this tapped powder will also be present in the CIPed composite sample. This is not the case with a CIPed composite sample using coarse W and fine Al (density 7.1 g/cm$^3$). Here the volume fraction of metal particles is significantly smaller (0.425) than the volume fraction of the granules (0.69) in the tapped mixture of coarse W and fine Al powders. This means that PTFE matrix is dispersing metal particles preventing them from forming force chains.

Based on this comparison it is reasonable to assume that, given the same constituent mass fractions, the strength of a composite will be higher when the volume fraction of the metallic particles in it are comparable to the corresponding value in tapped powders. This volume fraction can be significantly smaller than the volume fraction in a random packing of spherical particles (0.64) and it should depend on the shape of granules.

**Numerical modeling of shock wave in granular composite**

In the previous calculations the dynamic behavior of the samples were investigated under impact conditions but without shock wave propagating in the sample. For larger impact velocities the propagation of shock wave will result in new features depending on the particle size. For example, if particles of different strength and size are present in a mixture the distribution of internal energy depends on relative strength of particles and on their relative sizes.[16,17] In the mixture of soft and rigid, heavy particles the latter transform their kinetic energy, which they had before the shock front, into the thermal energy of small, soft particles (or soft matrix) behind the shock front. Thus, under shock loading, this mixture naturally transforms the kinetic energy of one component (large, rigid, heavy particles) into the internal energy of the other (small, soft, light particles or matrix). This results in a dependence of the distribution of internal energy between components on the size of the rigid, heavy particles.

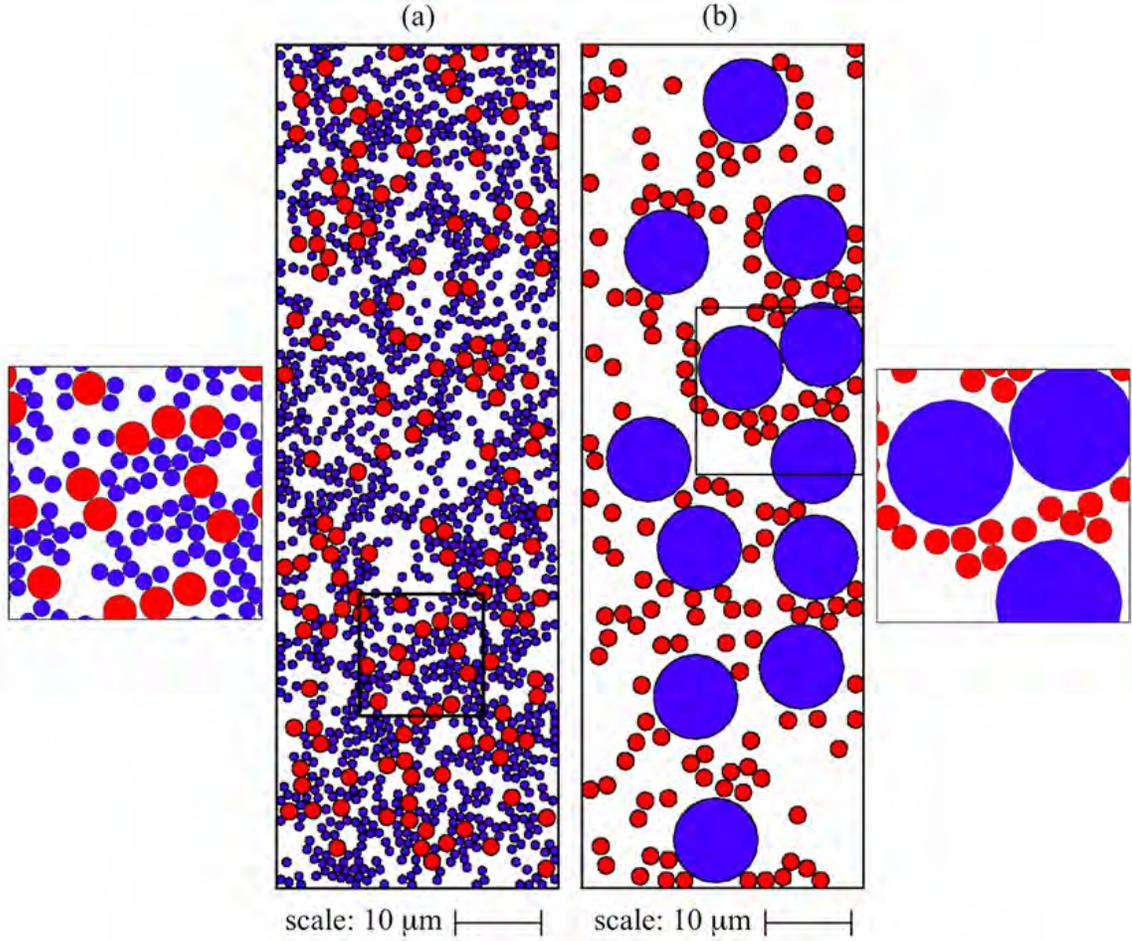

FIG. 11: (color online) Initial configuration of the shocked composite samples ($v_0 = 500$ m/s). (a) The sample with small W particles (diameter of 1 μm). The subfigure on the left shows a more detailed view of the microstructure. (b) The sample with large W particles (diameter of 10 μm). The subfigure on the right shows a more detailed view of the microstructure. The Al particles have a 2 μm diameter in both configurations.

Here, we investigate the redistribution of internal energy in the system where PTFE and Al particles represent a "soft" component and W particles represent hard and heavy component with relatively large particle sizes of W. For these calculations the size of the sample (i.e. number of particles) was increased to ensure that stationary states were achieved behind shock front in each sample at the end of calculation.

Two samples were prepared for numerical calculations with an impacting plate moving at a constant speed of 500 m/s. These samples have similar microstructures to those shown in Fig. 6 except that they are more than three times longer in the direction of the moving boundary such that the properties behind the shock waves may be investigated. The overall sizes of the samples are identical, but in Fig. 11 (a) the size of the W particles in the matrix are 1 μm and 10 μm in Fig. 11 (b). In the numerical calculations periodic boundary conditions were prescribed for the sides of the samples and a transmitting boundary condition was prescribed along the bottom of each sample to reduce the reflection of the shock from this boundary. Portions of the samples are outlined and detailed to the left of Fig. 11 (a) and to the right of Fig. 11 (b).

Time dependence of the ratios of internal energies for each component in the sample to the total internal energy are presented in Fig. 12 and the ratios of thermal energy increase for each component with respect to total internal energies at the end of calculations in the Table VI. The internal energy distribution is approximately constant after some initial period reflecting the establishment of an approximately stationary shock wave in calculations. The period of nonstationary wave propagation is extended in the system with large W particles due to their larger inertia, especially in relation to the internal energy of PTFE. It should be mentioned that the boundary conditions provided the energy flow from the impacting plate and was higher in the sample with small W particles due to its greater resistance to deformation. For this reason in our analysis below we use fractions of energies for each component with respect to total internal energies.

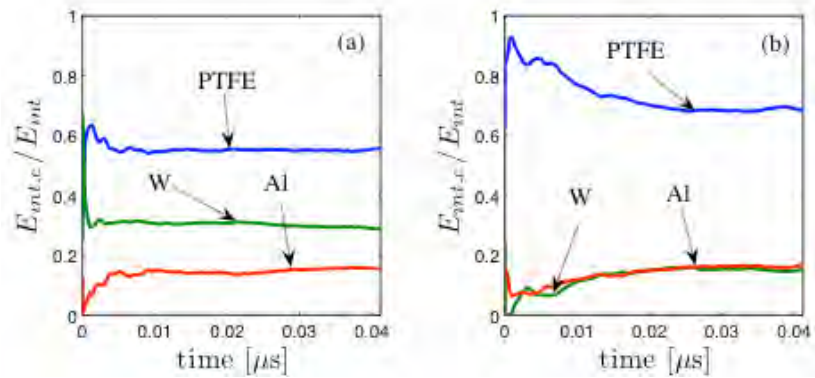

FIG. 12: (a) Fraction of internal energies in each material during the propagation of shock wave for the sample with small W particles. (b) Fraction of internal energies in each material during the propagation of shock wave the sample with large W particles.

TABLE VI. The mass ratio and fractions of increase of thermal energy ($E_{th}$) with respect to increase of total internal energy ($E_{int}$) of composite in samples with large and small W particles.

|      | Mass Ratio | $E_{th}/E_{int}$ | |
|------|------------|------------------|------------------|
|      | -          | Sample with small W | Sample with large W |
| PTFE | 0.23       | 0.18             | 0.28             |
| W    | 0.71       | 0.26             | 0.13             |
| Al   | 0.06       | 0.08             | 0.10             |

We can see that redistribution of the internal and thermal energy between components does not follow their mass fraction (0.23, 0.71, and 0.06, correspondingly for PTFE, W, and Al). This means that rigid W particles transform their kinetic energy mainly into the internal energy of PTFE particles and Al particles. This energy redistribution is more pronounced in the system with large W particles.

In Fig. 13 the temperature distribution derived from internal energy distribution is presented for two samples loaded in a similar way. Agglomerated small W and Al particles deform as groups and high temperature regions in the PTFE matrix occur between these groups (see the light gray coloring between these groups in the subfigure on the left of part Fig. 13 (a)). We can see that large W particles are less deformed (Fig. 13 (b)) than small W particles (Fig. 13 (a)). This translates into larger internal energy of PTFE and Al.

In the sample with large W particles high velocity 'jets' (the jets are indicated by arrows in the subfigure to the right of part Fig. 13 (b)) of the PTFE matrix flow between the large W particles and the Al particles. The latter moving between W particles are heavily deformed probably causing the increase of fraction of thermal energy deposited in Al particles (Table VI). In Fig. 13 (c) an averaged velocity profile (in the horizontal direction) of the material starting from the bottom of the samples is shown corresponding to (a) and (b). Curve (1) corresponds to the sample with small W particles and it may be assumed that the material from 20-80 μm above the bottom is stationary. Curve (2) corresponds to the sample with large W particles which demonstrates higher deviations from average value.

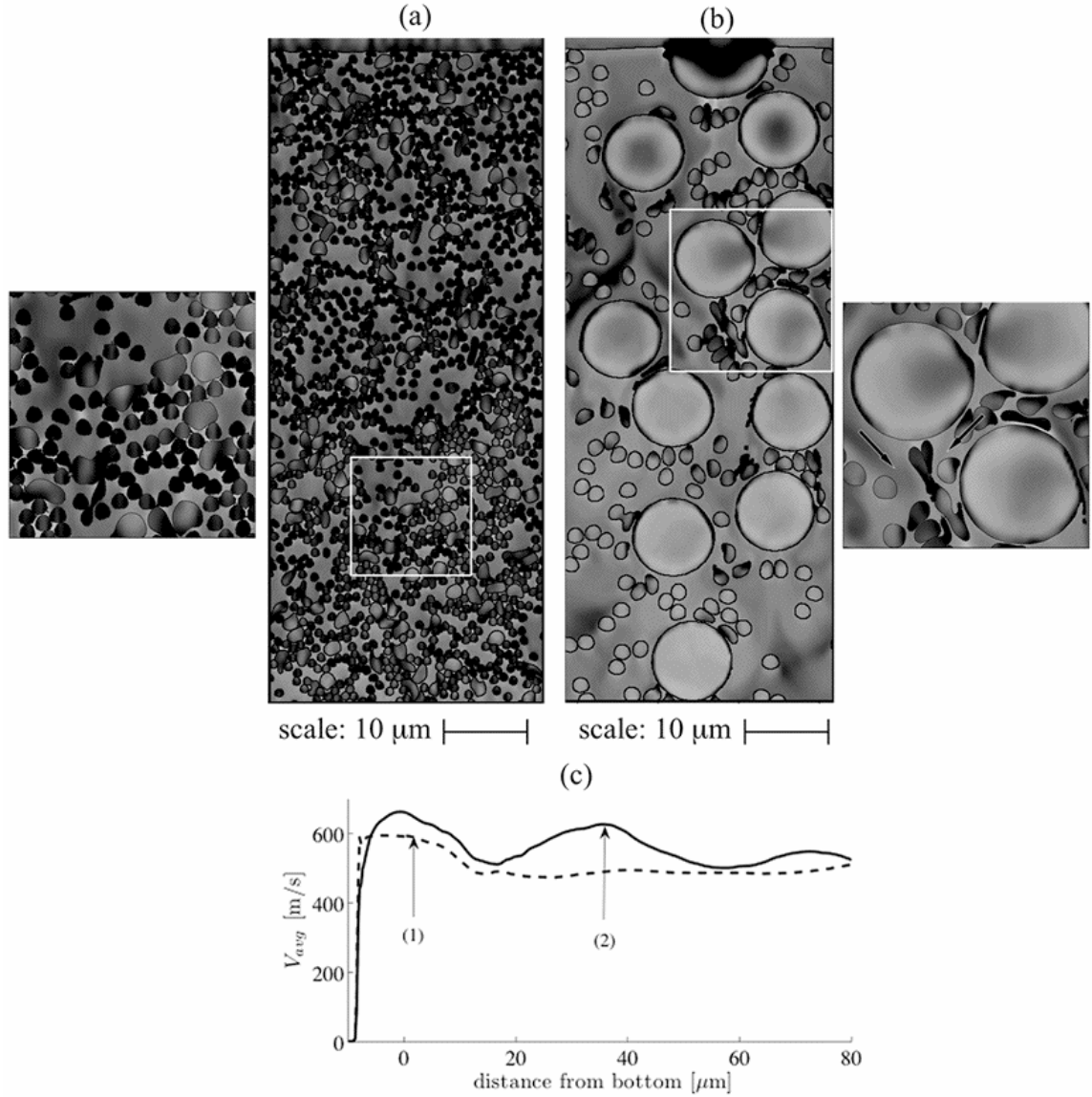

FIG. 13: Temperature distribution in shocked composite samples ($v_0 = 500$ m/s). The temperature scale ranges from 300 K (light gray) to $\geq 550$ K (black) for parts (a) and (b). (a) The sample with small W particles at 41.5 ns after impact. (b) The sample with large W particles at 38.3 ns after impact. (c) Averaged velocity profile (in the horizontal direction) of the material starting from the bottom of the samples shown in parts (a) and (b). Curve (1) corresponds to the sample with small W particles. Curve (2) corresponds to the sample with large W particles.

The preferential deposition of internal energy and specifically thermal energy in PTFE and Al can be very important for increase of temperature of PTFE and Al and their subsequent chemical interaction. We should notice that increase of the size of W particles resulted in significant increase of fraction of thermal energy of PTFE with some increase of thermal energy of Al particles. Large W particles facilitate localized flow in PTFE matrix. This results in relatively large displacements of Al particles in comparison with large W particles, their shear deformation and coalescence (see insert to Fig. 13 (c)), which may be conducive to their chemical interaction with PTFE. This behavior was absent in the granular composite with small W particles (compare inserts to Fig. 13 (a) and to Fig. 13 (c)).

The large difference in the temperature between different regions in shock compacted sample (Fig. 13 (a) and (b)) and relatively small variations of particle velocity (Fig. 13 (c)) reflects the existence of two times scales for equilibration behind shock wave – fast scale for establishing equilibration for particle velocity and slow scale for temperature equilibrium.

**CONCLUSIONS**

It was demonstrated that the dynamic mechanical properties of high-density mixtures of PTFE, Al and W powders could be tailored by changing the size of the particles and porosity of the mixture. A relatively weak polymer matrix allows the strength and fracture mode of this material to be governed by the granular type behavior of collections of metal particles. Composites with fine metallic particles and a higher porosity exhibited an unusually higher ultimate compressive strength than less porous composites having equivalent mass ratios with coarse W particles. The mesoscale force chains between the fine metallic particles are responsible for this unusual phenomenon as was demonstrated by the numerical calculations. Macrocracks were observed below the critical failure strain for the matrix and a competition between densification and fracture in some porous samples in dynamic tests.

Shock loading of this granular composite resulted in internal energy redistribution between components which can be tailored by a particles size of rigid component. It resulted in higher fraction of thermal energy of soft and light components (PTFE and Al) due to increase of size of heavy, rigid component (W particles). This may be important for ignition of reaction under shock loading in this (PTFE/Al) and similar systems.

**ACKNOWLEDGEMENTS**

The support for this project provided by ONR (N00014-06-1-0263 and MURI ONR Award N00014-07-1-0740) and the EPSRC (J.W. Addiss) is highly appreciated.